\documentclass[aps,prd,twocolumn,eqsecnum,showpacs]{revtex4}
\usepackage[dvips]{color,graphicx}
\usepackage{subfigure}
\usepackage{amsfonts}
\usepackage{amssymb}

\newcommand{\dalm}{\kern1pt\vbox{\hrule height 0.9pt\hbox{\vrule width
0.9pt\hskip 2.5pt\vbox{\vskip 5.5pt}\hskip 3pt\vrule width 0.3pt}\hrule height
0.3pt}\kern1pt}

\begin{document}


\title{
Self-similar cosmological solutions with dark energy. I:\\
formulation and asymptotic analysis
}

\author{
$^{1}$Tomohiro Harada\footnote{Electronic
address:harada@rikkyo.ac.jp},
$^{1,2,3,4}$Hideki Maeda\footnote{Electronic
address:hideki@cecs.cl} and 
$^{5,6}$B.~J.~Carr\footnote{Electronic address:B.J.Carr@qmul.ac.uk}}
\affiliation{
$^{1}$Department of Physics, Rikkyo University, Toshima, Tokyo 171-8501, Japan\\
$^{2}$Centro de Estudios Cient\'{\i}ficos (CECS), Arturo Prat 514, Valdivia, Chile\\
$^{3}$Department of Physics, International Christian University, 3-10-2 Osawa, Mitaka-shi, Tokyo 181-8585, Japan\\
$^{4}$Graduate School of Science and Engineering, Waseda University, Tokyo 169-8555, Japan\\
$^{5}$Astronomy Unit, Queen Mary, University of London, Mile End Road, London E1 4NS, UK\\
$^{6}$Research Center for the Early Universe, Graduate School of Science, University of Tokyo, Tokyo 113-0033, Japan}
\date{\today}

\begin{abstract}
Based on the asymptotic analysis of ordinary differential equations,
we classify all spherically symmetric self-similar solutions to the 
Einstein equations which are asymptotically 
Friedmann at large distances and contain a perfect fluid with equation of state $p=(\gamma -1)\mu$ with $0<\gamma<2/3$. This corresponds to a ``dark energy'' fluid and 
the Friedmann solution is accelerated in this
case due to anti-gravity.
This extends the previous analysis of spherically symmetric self-similar solutions for fluids with positive pressure ($\gamma>1$). 
However, in the latter case there is an 
additional parameter associated with the weak discontinuity 
at the sonic point and the solutions are only asymptotically
 ``quasi-Friedmann'', in the sense that they exhibit an angle deficit at large
 distances. In the $0<\gamma<2/3$ case, there is no sonic point
 and there exists a one-parameter family of solutions which are {\it genuinely}
asymptotically Friedmann at large distances. We find eight 
classes of asymptotic behavior: Friedmann or quasi-Friedmann or quasi-static or constant-velocity at large distances, quasi-Friedmann  or positive-mass singular or negative-mass singular at small distances, and quasi-Kantowski-Sachs at intermediate distances.
The self-similar
asymptotically quasi-static and quasi-Kantowski-Sachs solutions are analytically 
extendible and of great cosmological interest. 
We also investigate their conformal diagrams. 
The results of the present analysis are utilized in an accompanying paper to 
obtain and physically interpret numerical solutions.
\end{abstract}


\maketitle

\section{Introduction} 
There is great interest in spherically symmetric self-similar solutions to Einstein's equations because of their numerous applications in astrophysics and cosmology~\cite{cc1999}. 
Indeed there is now considerable evidence for the ``similarity hypothesis``, which postulates that spherically symmetric solutions may naturally evolve to self-similar form in a variety of situations.
The status of this hypothesis has been recently reviewed by Carr and Coley~\cite{cc2005}. 
In view of this, it is clearly important to have as complete a classification of spherically symmetric self-similar solutions as possible. 

Spherically symmetric self-similar solutions have the feature that all dimensionless quantities can be expressed in terms of $z=r/t$, where $r$ and $t$ are suitably chosen radial and time coordinates. 
If the source of the gravitational field is a single perfect fluid,
Cahill and Taub~\cite{ct1971} have shown that the only barotropic
equation of state compatible with the similarity assumption has the form
$p =(\gamma-1) \mu$ for some constant $\gamma$, where $\mu$ and $p$ are
the energy density and the pressure, respectively.  
In this case, for a given value of $\gamma$, the solutions are described by two parameters and, providing one restricts attention to shock-free perfect fluids with positive pressure ($1<\gamma \le 2$), they have been completely classified.
This classification has been achieved using a combination of the ``comoving'' approach (in which the coordinates are adapted to the fluid 4-velocity) and the ``homothetic'' approach (in which the coordinates are adapted to the homothetic vector). 
These approaches have been used by Carr and Coley~\cite{cc2000a} (henceforth CC) and Goliath et al.~\cite{gnu}, respectively, although a full understanding of the solutions requires that one combines them~\cite{ccgnu1}. 

A key step in the CC analysis is the derivation of all possible asymptotic behaviors at large and small distances~\cite{cc2000b}. 
For positive pressure, there are at least three kinds of behavior at large spatial distances (usually corresponding to the limit $z \to \infty$): (1) asymptotically Friedmann (1-parameter); (2) asymptotically Kantowski-Sachs (1-parameter), though these are probably unphysical for  $\gamma > 2/3$); and (3) asymptotically quasi-static (2-parameter). 
There are also two families of solutions which exist only when $\gamma > 6/5$: (4) asymptotically Minkowski at infinite $z$ (1-parameter); and (5) asymptotically Minkowski at finite $z$ but infinite physical distance (2-parameter). 
At small spatial distances, the solutions are of four kinds: they contain either (a) a black hole singularity or (b) a naked singularity at finite $z$ (but zero physical distance) or they can be connected to $z=0$ via a sonic point, in which case they are either (c) static or (d) represent a perturbation of the Friedmann solution. 

The complete family of $1<\gamma \le 2$ solutions can now be found by combining the five kinds of large-distance behavior and four kinds of small-distance behavior. 
The way in which one connects the large-distance and small-distance solutions depends crucially on whether or not the solution passes through a sonic point. 
If the solutions remain supersonic (or subsonic) everywhere, then the small-$z$ behavior is uniquely determined by the large-$z$ behavior. 
However, if there is a sonic point, the behavior of the solutions is much more complicated because the equations do not determine their behavior uniquely there, so there can be a discontinuity. 
Indeed, only a subset of solutions are ``regular''  at a sonic point in
the sense that the pressure gradient is finite and they can be extended
beyond it. Because of this feature, the family
of solutions with a regular center and a regular sonic point
have a band structure, 
with the solutions which are analytic at the center 
and sonic point forming a discrete subset of these~\cite{op1987,op1990,fh1993}.
This feature is very important for naked singularity formation 
and critical behavior in the
gravitational collapse of a perfect fluid~\cite{hm2002,ec1994}.

In this and the accompanying papers~\cite{mhc2} (henceforth Paper II), we take a first step in extending the classification of CC to the negative pressure case ($\gamma <1$), using a combination of numerical and analytical studies. 
In fact, part of the work is already done, because ref.~\cite{cc2000b}
does include the asymptotic analysis for $\gamma < 1$. 
However, there are some errors in that work (see Appendix A) and  the full family of solutions has not yet been analyzed. 
This paper will focus particularly on the case with $0<\gamma<2/3$. 
Although this equation of state violates the strong energy condition and was little emphasized by CC, it may be very relevant for cosmology -- both in the early universe (when inflation occurred~\cite{linde}) and at the current epoch~\cite{supernova} (when acceleration may be driven by some form of ``dark energy''). 
In both situations the matter model exhibits ``anti-gravity'' 
but the dominant, null and weak energy conditions still hold.

As in the positive pressure case, it should be stressed that our classification does not cover imperfect or multiple fluids or solutions with shocks.
The accelerated expansion of the universe can also be realized by a scalar field with a flat potential (i.e.  quintessence). 
The non-existence of self-similar black hole solutions embedded in an
exact Friedmann universe for this case was already demonstrated in ref.~\cite{hmc2006} and this is clearly complementary to the present work.

There are several key differences between the $0<\gamma<2/3$ and $2/3<\gamma<2$ cases.
A purely formal difference is that the limiting values of $z$ for large and small spatial distances in the flat Friedmann solution are reversed: large spatial distances now correspond to $z \rightarrow 0$ and small ones to $z \rightarrow \infty$. 
The other differences have more physical significance. 
The first is that there are no sound-waves, since the sound-speed $c_s = \sqrt{dp/d\mu} = c \sqrt{\gamma -1}$ is imaginary.
This considerably simplifies the analysis, since there can be no discontinuities and solutions are analytic everywhere. 
The second is that there is no exact static solution. 
although there are still asymptotically quasi-static solutions (a point which was missed in ref.~\cite{cc2000b}). 
On the other hand, the Kantowski-Sachs and asymptotically quasi-Kantowski-Sachs solutions now become physical. 
Finally, the solutions which were asymptotically Minkowski at large distances are now replaced with solutions which are asymptotically singular at small distances. 

Despite these differences, many features of the CC classification still apply. 
In particular, there is still a 1-parameter family of solutions asymptotic to the flat Friedmann model at large distances 
and these are of great physical interest since they may have cosmological applications. 
In the positive-pressure case, some of these solutions are supersonic everywhere and contain black holes which grow at the same rate as the particle horizon~\cite{ch,bh1}.  
Others represent density perturbations in a Friedmann background which always maintain the same form relative to the particle horizon~\cite{cy}. 
However, recently it was pointed out  that none of these positive-pressure solutions are ``properly'' asymptotic Friedmann
because they exhibit a solid angle deficit at infinity~\cite{mkm2002}. 
They may still be relevant to the real universe (since observations may not preclude such an angle deficit) but it would be more accurate to describe them as ``quasi-Friedmann''.

In the $0<\gamma<2/3$ case, we will show that there are genuine asymptotically Friedmann solutions.
We will analyze these solutions numerically and
exploit these results in Paper II~\cite{mhc2} to interpret the solutions physically .
The key feature is the existence of 
asymptotically Kantowski-Sachs and static solutions, 
both of which are extendible analytically. 
This leads to the possibility of cosmological black hole,  
wormhole and white hole solutions.

The plan of this paper is as follows. In Section II, 
we present the 
basic field equations 
for spherically symmetric self-similar spacetimes.
We describe some exact solutions in Section III and 
solutions which are asymptotic to these In Section IV. We summarize our results and discuss their implications in Section V.

\section{Basic equations} 
We consider a spherically symmetric spacetime with the line element
\begin{equation}
\label{metric_original}
ds^2=-e^{2\Phi(t,r)}dt^2+e^{2\Psi(t,r)}dr^2+R(t,r)^2d\Omega^2,
\end{equation}  
where $d\Omega^2 \equiv d\theta^2+\sin^2 \theta d\varphi^2$. We take the matter field to be a perfect fluid with energy-momentum tensor 
\begin{eqnarray}
T_{\mu\nu} = p g_{\mu\nu}+(\mu+p) u_{\mu} u_{\nu},
\end{eqnarray}
where $p$ and $\mu$ are the pressure and the energy density, respectively, $u_{\mu}$ is the $4$-velocity of the fluid, and we use units with $c=1$. 
We will adopt comoving coordinates, so that the $4$-velocity is 
\begin{eqnarray}
u^\mu\frac{\partial}{\partial x^\mu}=e^{-\Phi}\frac{\partial}{\partial t}.
\end{eqnarray}  
The field equations can then be written in the following form:
\begin{eqnarray}
p_{,r}&=& -(\mu+p)\Phi_{,r} \; , \label{basic1}\\
\mu_{,t}&=& -(\mu+p)\left(\Psi_{,t}+2\frac{R_{,t}}{R}\right), \label{basic2}\\
m_{,r} &=& 4\pi\mu R_{,r} R^2,   \label{basic3}\\
m_{,t} &=& -4\pi p R_{,t} R^2, \label{basic4}\\ 
0&=&-R_{,rt}+\Phi_{,r} R_{,t}+\Psi_{,t}R_{,r},\label{basic5}\\
m &=& \frac{R}{2G}(1+e^{-2\Phi}{R_{,t}}^2-e^{-2\Psi}{R_{,r}}^2),\label{basic6}
\end{eqnarray}  
where a comma denotes partial differentiation and $G$ is the
gravitational constant. 
The first two equations correspond to the energy-momentum conservation and the next two specify the Misner-Sharp mass $m$. 
Five of the above six equations are independent. 
Throughout this paper, we call the direction of increasing
(decreasing) $t$ the future (past) direction.

A spacetime is
self-similar if it admits a homothetic Killing vector $\xi^\mu$, which is defined by
\begin{eqnarray}
{\cal{L}}_{\bf\xi} g_{\mu\nu} = 2g_{\mu\nu},
\end{eqnarray}
where ${\cal{L}}_{\bf\xi}$ denotes the Lie derivative along $\xi^\mu$.
Cahill and Taub~\cite{ct1971} first investigated spherically symmetric self-similar solutions in which the homothetic Killing vector is neither parallel nor orthogonal to the fluid flow vector. 
They showed that -- by a suitable coordinate transformation -- such solutions can be put into a form in which all dimensionless quantities are functions only of the dimensionless variable $z \equiv r/t$.
In this case, $\xi^\mu$ is given by
\begin{eqnarray}
\xi^\mu\frac{\partial}{\partial x^\mu}=t\frac{\partial}{\partial t}+r\frac{\partial}{\partial r}\;.
\end{eqnarray}  

The line element in a spherically symmetric self-similar spacetime can be written as
\begin{equation}
ds^2 = -e^{2\Phi(z)}dt^2+e^{2\Psi(z)}dr^2+ r^2 S^2(z) d\Omega^2. \label{metric}
\end{equation}
A hypersurface $\Sigma$ 
of constant $z$ is called a similarity surface and is generated by the homothetic Killing vector.
The induced metric on $\Sigma$ is 
\begin{equation}
ds^{2}_{\Sigma}=-(1-V^{2})e^{2\Phi}dt^{2}+r^{2}S^{2}d\Omega^{2}, \label{metric2}
\end{equation}
where $V\equiv |z|e^{\Psi-\Phi}$ is the speed of the fluid relative to the
surfaces of constant $z$.
The area radius $R=rS$ must be positive but 
$S$ and $r$ need not be. (Note that CC adopt a convention in which $r$ is always positive.)
A similarity surface is spacelike for $(V^2-1)e^{2\Phi}>0$, 
timelike for $(V^2-1)e^{2\Phi}<0$ and null for $(1-V^{2})e^{2\Phi}=0$. In the latter case, $\xi^{\mu}$ is also null and the similarity surface is called a 
similarity horizon. 

It is helpful 
to rewrite Eq.~(\ref{metric}) in the conformally static form:
\begin{eqnarray}
ds^{2}&=&-e^{2\tau}\left[e^{2\Phi}\left\{(1+V)d\tau
 +\frac{V}{z}dz\right\}\left\{(1-V)d\tau-\frac{V}{z}dz\right\}\right. \nonumber \\
&&\left.+z^{2}S^{2}d\Omega^{2}\right],
\end{eqnarray}
where the conformal factor depends only on $\tau \equiv \ln |t|$. 
To obtain the causal structure of the self-similar spacetime,
we can then use the analogy 
with the static spacetime. This is because the homothetic Killing vector, 
similarity surfaces and similarity horizons 
are the counterparts of the Killing vector, Killing surfaces and
Killing horizons in the static case
(see e.g. ref.~\cite{cg2003}). 

The Einstein equations imply that $p$, $\mu$ and $m$ must have the form 
\begin{eqnarray}
8\pi G\mu&=&\frac{W(z)}{r^{2}}, \label{defW}\\
8\pi Gp&=&\frac{P(z)}{r^{2}},\label{defP} \\
2Gm&=&rM(z),\label{defM}
\end{eqnarray}
where we assume that the energy density is non-negative ($W\ge 0$).
Then the Einstein and energy-momentum conservation
equations reduce to ordinary differential equations for the
non-dimensional functions with respect to the self-similar variable $z$:
\begin{eqnarray}
-2P+P'&=& -(W+P)\Phi', \label{basic1z}\\
-W'&=& (W+P)\left(\Psi'+2\frac{S'}{S}\right), \label{basic2z}\\
M+M'&=& WS^2(S+S'),   \label{basic3z}\\
-M' &=& PS^2S', \label{basic4z}\\ 
0&=&S''+S'-\Phi'S'-\Psi'(S+S'),\label{basic5z}\\
M &=& S[1+z^2e^{-2\Phi}{S'}^2-e^{-2\Psi}(S+S')^2],\label{basic6z}
\end{eqnarray}  
where a prime denotes a derivative with respect to $\ln |z|$.

We assume the equation of state has the form $p = (\gamma-1)\mu$, which is the only barotropic one compatible with the homothetic assumption.
The dominant energy condition requires 
$0 \le \gamma \le 2$.
In this paper we 
exclude the value of 1 (corresponding to dust) since this
needs special treatment and has been analyzed in ref.~\cite{carr2000}.
We also exclude $\gamma=0$, corresponding to a cosmological constant, since this is 
incompatible with self-similarity.
Equations~(\ref{basic1z}) and (\ref{basic2z}) can then be integrated to give
\begin{eqnarray} 
e^{\Phi}&=&c_0z^{2(\gamma -1)/\gamma}W^{-(\gamma-1)/\gamma},\label{integral1}\\ 
e^{\Psi}&=&c_1 S^{-2}W^{-1/\gamma},\label{integral2}
\end{eqnarray} 
where $c_0$ and $c_1$ are integration constants.
The velocity function $V$ can be shown to be
\begin{equation}
V =\frac{c_1}{c_0}
z^{(2-\gamma)/\gamma}S^{-2}W^{(\gamma-2)/\gamma}.\label{speed}
\end{equation}
 
Equations~(\ref{basic1z})--(\ref{basic6z}) reduce to ordinary
differential equations for $S$, $S'$ and $W$:
\begin{eqnarray}
&& \frac{V^2-(\gamma-1)}{\gamma}\frac{W'}{W}=\frac{\gamma
 c_1^2W^{(\gamma-2)/\gamma}}{2S^4}-\frac{2(\gamma-1)}{\gamma}\nonumber \\
&&\quad\quad~~~~~~~~~~~~~~~~ -2V^2\frac{S'}{S},\label{00+11}\\
&&S''=S'\left({\gamma-2 \over \gamma}-\frac{(\gamma-1)W'}{\gamma W}\right) \nonumber \\
&&\quad\quad-(S+S')\left(\frac{2S'}{S}+\frac{W'}{\gamma W}\right),\label{01}\\
&&M=WS^2(\gamma S'+S),\label{m1}\\
&&M=S\biggl[1+c_0^{-2}z^{2(2-\gamma)/\gamma}W^{2(\gamma-1)/\gamma}{S'}^2  \nonumber \\
&&\quad\quad -c_1^{-2} S^{4}W^{2/\gamma}(S+S')^2\biggl].
\label{m2}
\end{eqnarray}
From Eqs.~(\ref{m1}) and (\ref{m2}), we obtain the following relation between $S$, $S'$ and $W$:
\begin{eqnarray}
&& WS^2(\gamma S'+S) =S\biggl[1+c_0^{-2}z^{2(2-\gamma)/\gamma}W^{2(\gamma-1)/\gamma}{S'}^2
 \nonumber \\
&& \quad \quad~~~~~~~~~~~~~ -c_1^{-2} S^{4}W^{2/\gamma}(S+S')^2\biggl].
\label{eq:constraint}
\end{eqnarray}
Using this and Eq.~(\ref{m1}), we obtain another constraint between $M$, $S$ and $W$:
\begin{eqnarray}
&&\left(1-\frac{M}{WS^{3}}\right)^{2}V^{2}
-\left(\gamma-1+\frac{M}{WS^{3}}\right)^{2} \nonumber \\
&&\quad\quad  +\gamma^{2}c_{1}^{2}W^{-2/\gamma}S^{-6}\left(1-\frac{M}{S}\right)=0.
\end{eqnarray}

\section{Exact solutions}
\subsection{Friedmann solution}
The flat Friedmann solution in self-similar coordinates
corresponds to
\begin{equation}
S=S_{0}z^{-2/(3\gamma)},\quad W=W_{0}z^{2}, 
\label{eq:friedmann_S_W}
\end{equation}
where $S_{0}$ and $W_{0}$ are constants. $V$ is determined by 
Eq.~(\ref{speed}) as
\begin{equation}
V=V_{0}z^{1-2/(3\gamma)},\quad V_{0}=\frac{c_{1}}{c_{0}}S_{0}^{-2}
W_{0}^{(\gamma-2)/\gamma}.
\label{eq:V_Friedmann}
\end{equation}
and Eq.~(\ref{01}) holds trivially. From Eqs.~(\ref{00+11}) and (\ref{eq:constraint}),
 $S_{0}$ and $W_{0}$ must satisfy
\begin{eqnarray}
\frac{2}{3\gamma}V_{0}^{2}&=&\frac{\gamma
 c_{1}^{2}W_{0}^{(\gamma-2)/\gamma}}
{2 S_{0}^{4}}, 
\label{eq:relation_1_Friedmann}\\
1&=&c_{1}^{-2}S_{0}^{6}W_{0}^{2/\gamma}\left(1-\frac{2}{3\gamma}\right)^{2}.
\label{eq:relation2_Friedmann}
\end{eqnarray}
These equations are consistent provided $\gamma$ is not $2/3$ or $0$. 
Since $S_0$ and $W_0$ are determined in terms of $\gamma$, $c_0$ and $c_1$, where $c_0$ and $c_1$ are just gauge constants, there is no free parameter in this solution for a given value of  $\gamma$.

Introducing constants $a_{0}$ and $b_{0}$ in place of $S_{0}$ and
$W_{0}$, we obtain:
\begin{eqnarray}
e^{\Phi}&=&e^{\Phi_{\rm F}}\equiv a_0,\label{FRWsolution1}\\
e^{\Psi}&=&e^{\Psi_{\rm F}}\equiv b_0z^{-q},\\
S&=&S_{\rm F} \equiv \frac{b_0}{|1-q|}z^{-q}, 
\label{eq:friedmann_S}\\
W&=&W_{\rm F}\equiv \frac{4}{3\gamma^2a_0^2}z^{2}, 
\label{eq:friedmann_W}\\
M&=&M_{\rm F}\equiv \frac{4b_0^3}{9\gamma^2 a_0^2|1-q|^3}z^{2-3q}. \label{FRWsolution3}
\end{eqnarray}
Here the constants $a_0$ and $b_0$ can be chosen arbitrarily and 
\begin{equation}
q \equiv \frac{2}{3\gamma}. \label{q}
\end{equation}
The constants $c_{0}$ and $c_{1}$ in Eqs.~(\ref{integral1}) 
and (\ref{integral2}) are given in terms of $a_{0}$ and $b_{0}$ as
\begin{eqnarray}
c_0&=&a_0\left(\frac{4}{3\gamma^2 a_0^2}\right)^{(\gamma-1)/\gamma},\\
c_1&=&\frac{b_0^3}{(1-q)^2}\left(\frac{4}{3\gamma^2 a_0^2}\right)^{1/\gamma},
\end{eqnarray}
where Eqs.~(\ref{eq:friedmann_S_W})--(\ref{eq:relation2_Friedmann}),
(\ref{eq:friedmann_S}) and (\ref{eq:friedmann_W}) are used.
One can put the metric into a more familiar form, 
\begin{eqnarray}
ds^2=-d{\tilde t}^2+{\tilde t}^{2q}(d{\tilde r}^2+{\tilde r}^2 d\Omega^2),
\end{eqnarray}
by introducing new coordinates
\begin{equation} 
{\tilde t}=a_0t, \;~~~ {\tilde r}=a_{0}^{-q} b_0 r^{1-q}/|1-q|.
\end{equation}
It should be noted that ${\tilde r}=\infty$ corresponds to $r=\infty$
for the decelerating case ($2/3<\gamma\le 2$) but to $r=0$ for the
accelerating one ($0<\gamma<2/3$). In the $\gamma=2/3$ case, the 
homothetic Killing vector is parallel to the fluid flow and these equations no longer apply~\cite{mh2006b}.  

\begin{figure}[htbp]
\begin{center}
\includegraphics[width=0.6\linewidth]{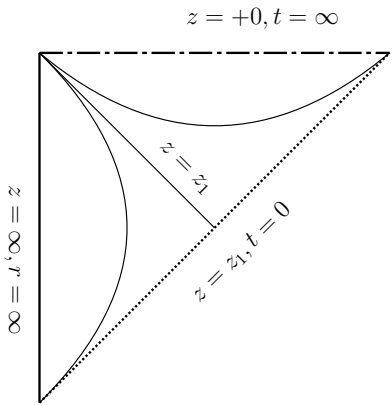}
\caption{\label{fig:Friedmann} The conformal diagram of the 
flat Friedmann solution for $0<\gamma<2/3$. There is a 
similarity horizon at $z=z_{1}$ ($>0$), 
corresponding to the cosmological event horizon,
and a regular center at $z=\infty$ and $r=\infty$.
 The initial singularity at $t=0$ is null, 
while null infinity is given by $t=\infty$.
The dotted and dot-dashed lines denote the singularity and 
infinity, respectively.
The thin solid curves and lines denote 
similarity surfaces, i.e. orbits of $z=$constant.
}
\end{center}
\end{figure}

There exists a finite non-zero $z_{1}$ 
where $V$ crosses 1.
Figure~\ref{fig:Friedmann} shows the conformal diagram of the 
flat Friedmann solution for $0<\gamma<2/3$, including the similarity 
surfaces. We can see that the initial singularity 
is null.
The similarity horizon $z=z_{1}$
corresponds to 
the cosmological event horizon, while 
$z=\infty$ and $r=\infty$ is a regular center.
The dotted and dot-dashed lines denote the singularity and 
infinity, respectively.

\subsection{Kantowski-Sachs solution}
The Kantowski-Sachs solution in self-similar coordinates 
corresponds to
\begin{equation}
S=S_{0}z^{-1}, \quad W=W_{0}z^{2}.
\end{equation}
Then Eq.~(\ref{01}) holds trivially and Eq.~(\ref{eq:constraint}) yields
\begin{equation}
(1-\gamma)W_{0}S_{0}^{3}=S_0[1+c_0^{-2}W_0^{2-2/\gamma}S_0^{2}].
\label{eq:AKS_MASS_CONSISTENCY}
\end{equation}
From Eq.~(\ref{speed}), we have
\begin{equation}
V=V_{0}z^{3-2/\gamma}, \quad V_{0}=\frac{c_{1}}{c_{0}}S_{0}^{-2}W_{0}^{1-2/\gamma},
\end{equation}
and hence $V\to 0$ as $z\to \infty$.
Eq.~(\ref{00+11}) is trivially satisfied 
to lowest order but determines $W_{0}$ to the next lowest order:
\begin{equation}
\left(\frac{1}{\gamma}-1\right)c_{0}^{-2}
=\frac{\gamma}{4}W_{0}^{-1+2/\gamma}.
\label{eq:AKS_W0}
\end{equation}
Equations~(\ref{eq:AKS_MASS_CONSISTENCY}) and (\ref{eq:AKS_W0})
then give
\begin{equation}
\frac{(2-3\gamma)(2-\gamma)}{4(1-\gamma)}W_{0}S_{0}^{2}=1,
\end{equation}
so we can obtain a physical solution only for $0<\gamma<2/3$ and 
there remains no free parameter in this case. This solution was first
obtained in ref.~\cite{ck}, although that paper introduced different
variables to deal with the unphysical solutions for the $2/3<\gamma<2$ case.

The Kantowski-Sachs solution in more standard coordinates 
can be written as~\cite{ks1966}
\begin{eqnarray}
ds^2&=&-\frac{(2-3\gamma)(2-\gamma)}{\gamma^2}dt^2+t^{4(1-\gamma)/\gamma}d\bar{r}^2 \nonumber \\
&&~~~~~~~~+t^2(d\theta^2+\sin^2\theta d\phi^2), \\
8\pi G\mu&=&\frac{4(1-\gamma)}{(2-3\gamma)(2-\gamma)t^2},\\
2Gm&=&\frac{4(1-\gamma)^2t}{(2-3\gamma)(2-\gamma)},\\
V&=&\frac{|\gamma|z^{-(2-3\gamma)/\gamma}}{\sqrt{(2-3\gamma)(2-\gamma)}},
\end{eqnarray}
where $\bar{r}$ is a radial coordinate. 
As expected, $V$ is only real for $0<\gamma<2/3$. 
In the Kantowski-Sachs solution, the area of the 2-sphere with constant $t$ and $r$ does not depend on $r$ but only on $t$, so
it expands with time. The topology of the constant $t$ spacelike
hypersurface is $S^{2} \times R$.

\begin{figure}[htbp]
\begin{center}
\includegraphics[width=0.8\linewidth]{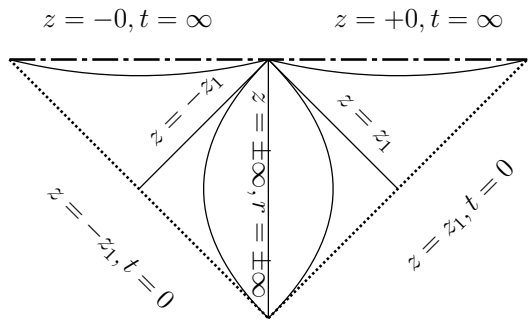}
\caption{\label{fig:KS} The conformal diagram of the 
Kantowski-Sachs solution for $0<\gamma<2/3$. The solution 
is analytically extendible beyond $z=\pm\infty$ and $r=\pm \infty$.
There are two similarity horizons $z=\pm z_{1}$, 
both corresponding to cosmological event horizons.
The initial singularity at $t=0$ is null 
and future null infinity is 
given by $t=\infty$.}
\end{center}
\end{figure}

There again exists a non-zero finite $z_{1}$
 where $V$ crosses 1.
$r=\infty$ or $z=\infty$ for fixed $t(>0)$ 
can be analytically extended to negative $r$ or negative $z$
beyond $z=\pm\infty$.
Figure~\ref{fig:KS} shows the conformal diagram of the (extended)
Kantowski-Sachs solution for $0<\gamma<2/3$. The similarity 
surfaces are also shown. We can see that the initial singularity 
is null.
There are two similarity horizons $z=\pm z_{1}$, corresponding to 
two cosmological event horizons. The corresponding 
conformal diagram (Fig. 15) in ref.~\cite{cg2003} is incorrect. 

\subsection{Absence of static solution}
The static solution would need to have 
\[S=S_{0}, \quad W=W_{0},\]
in which case Eq.~(\ref{speed}) gives
\begin{equation}
V=V_{0}z^{(2-\gamma)/\gamma}, \quad V_{0}=\frac{c_{1}}{c_{0}}S_{0}^{-2}W_{0}^{(\gamma-2)/\gamma}.
\end{equation}
Equations~(\ref{00+11}) and (\ref{eq:constraint}) then reduce to
\begin{equation}
0=\frac{\gamma
 c_{1}^{2}W_{0}^{(\gamma-2)/\gamma}}{S_{0}^{4}}+2\left( \frac{1-\gamma}{\gamma}\right)
 \label{eq:ST_00+11_CONSISTENCY}
\end{equation}
and 
\begin{equation}
W_{0}S_{0}^{3}=S_{0}[1-c_{1}^{-2}S_{0}W_{0}^{2/\gamma}],
\label{eq:ST_MASS_CONSISTENCY}
\end{equation}
respectively. For $\gamma>1$, Eq.~(\ref{eq:ST_00+11_CONSISTENCY}) can 
be satisfied and there is a static self-similar solution, 
but there is no such solution for $0\le \gamma< 1$
However, as we will see later, the absence of an exact static solution 
does not preclude the possibility of
an asymptotically static solution.

\section{Asymptotic behaviors}

\subsection{Friedmann asymptote for $z\to 0$}

We now focus on self-similar solutions which are asymptotic to the flat Friedmann model at large spatial distances, i.e. in which $\Phi$, $\Psi$, $S$, $W$ and $M$ approach the form given by Eqs.~(\ref{FRWsolution1})-(\ref{FRWsolution3}) as ${\tilde r} \rightarrow \infty$.
In this case, it is convenient to introduce new functions $A$ and $B$,
defined by
\begin{eqnarray}
W&=&W_{\rm F}(z) e^{A(z)},\label{defA}\\
S&=&S_{\rm F}(z) e^{B(z)}, \label{defBC}
\end{eqnarray}
which describe the deviations from the flat Friedmann solution. 
Equations~(\ref{00+11}), (\ref{01}) and (\ref{speed}) can then be written as
\begin{eqnarray}
A'&=&\frac{\gamma V^2[3\gamma q^2 e^{-(\gamma-2)A/\gamma}-2(q+2B')]}{2[V^2-(\gamma-1)]}, \label{fulleq1} \\
B''&=&-(B'-q)\left(1-q+B'+\frac{\gamma-1}{\gamma}A'\right) \nonumber \\
&&-\left(q+2B'+\frac{1}{\gamma}A'\right)(1-q+B'), \label{fulleq2} \\
V&=&\frac{b_0}{a_0}z^{1-q}e^{-2B+(1-2/\gamma)A}, \label{const}\label{fulleq3}
\end{eqnarray}
respectively.
Equation (\ref{eq:constraint}) yields another relation:
\begin{eqnarray}
&&\frac{4b_0^3}{9\gamma^2 a_0^2|1-q|^3}
z^{2-3q}e^{A+3B}(1+3\gamma B')\nonumber \\
&=&\frac{b_0}{|1-q|}z^{-q}e^B\biggl[1+\frac{b_0^2}{a_0^2(1-q)^2}z^{2-2q} \nonumber \\
&&~~\times e^{2(\gamma-1)A/\gamma+2B}(-q+B')^2 \nonumber \\
&&~~-\frac{1}{(1-q)^2}e^{2A/\gamma+6B}(1-q+B')^2\biggl].
\end{eqnarray}

The flat Friedmann solution is given by $A=B=0$ for all $z$.
Spacelike infinity corresponds to $z \to \infty$ for the 
decelerating case ($2/3<\gamma\le 2$) and $z \to 0$ for the 
accelerating case ($0<\gamma<2/3$).
Note that $V\propto z^{1-2/(3\gamma)}\to\infty$ at large distances
if $A$ and $B$ are finite.
This means that $z=\infty$ ($2/3<\gamma\le 2$) and 
$z=0$ ($0<\gamma<2/3$)
are horizontal lines in a conformal diagram.

The asymptotic form of self-similar solutions which approach the flat
Friedmann at large distances can be found by neglecting the $V^{-2}$
term and linearizing the equations with respect to $A$, $A'$, $B'$ and
$B''$. (It is noted that $B$ does not need to be small.)
From Eqs.~(\ref{fulleq1}) and (\ref{fulleq2}), the linearized equations
at spacelike infinity become
\begin{eqnarray}
0&=&3q^2(\gamma-2)A+4B'+\frac{2}{\gamma}A',   \\
0&=&B''+3(1-q)B'+\left( \frac{1-\gamma q}{\gamma}\right) A'
\end{eqnarray}
in both cases.
These equations lead to two independent solutions and
the general solution is given by a linear combination of these.

The first solution is
\begin{eqnarray}
A(z)&= &A_1 z^{2(2-3\gamma)/(3\gamma)}+\cdots, \\
B(z)&=&\beta+B_1 z^{2(2-3\gamma)/(3\gamma)}+\cdots,
\end{eqnarray}
where
\begin{eqnarray}
A_1&\equiv&(2-3\gamma)B_1,\\
B_1&\equiv&-\frac{{a_0}^2}{{b_0}^2}\frac{\gamma(3\gamma-2)}{4(3\gamma+2)}(e^{-2\beta}-e^{4\beta}).
\end{eqnarray}
$\beta$ is a constant and the dots denote higher order terms.
$A$ and $B$ converge at large distances in both the decelerating ($2/3<\gamma \le 2$) 
and accelerating ($0<\gamma<2/3$) cases. 
This solution was first obtained in ref.~\cite{ch} for the radiation case ($\gamma =4/3$) and in ref.~\cite{cy} for more general $\gamma$.
However, it is not properly asymptotic to the flat Friedmann solution
because Eq.~(\ref{defBC}) shows that there is a residual $e^{\beta}$ term
in $S$ at infinity, so Eq.~(\ref{metric}) implies that there is a
solid angle deficit~\cite{mkm2002}.
We describe solutions with this asymptotic behavior
as {\it asymptotically quasi-Friedmann}. 

The second solution is 
\begin{eqnarray}
A&\approx & A_0 z^{(2-\gamma)/\gamma}, \label{Atrue}\\
B&\approx & -\frac{1}{6\gamma}A_0 z^{(2-\gamma)/\gamma},\label{Btrue}
\end{eqnarray}
where $A_0$ is a constant and $x\approx y$
means $(x/y) \rightarrow 1$ in the relevant limit.
The above solution was also found in ref.~\cite{nusser2006}.  
The decelerating solution must be discarded, since this diverges at large distances ($z \to \infty$).   
On the other hand, the accelerating solution 
is ``properly'' asymptotic to Friedmann since $A$ and $B$ both converge to
zero at large distances ($z \to 0$), so there is no solid angle
deficit. In the following, we generally omit the term ``properly'' 
and simply describe these solutions as {\it asymptotically Friedmann}.

We conclude that there is a 1-parameter family of 
asymptotically Friedmann solutions at large distances for
$0<\gamma<2/3$. Note that, in the positive pressure case ($\gamma >1$), 
it is well known that there
is no physical solution which is {\it exactly} Friedmann at sufficiently large
but finite distances~\cite{ch}. Although one might envisage attaching an
interior black hole solution to an exterior Friedmann solution at a sonic
point (since there can be a discontinuity there), no such solution is
possible. This conclusion trivially applies when the pressure is
negative ($\gamma <1$) because there are no sound-waves in this case. 

\subsection{Friedmann asymptote for $z\to \infty$}
We are also interested in solutions which are asymptotically Friedmann at {\it small} distances from the origin. We therefore seek solutions which have 
\begin{equation}
S\approx S_{0}z^{-2/(3\gamma)},\quad W\approx W_{0}z^{2}
\end{equation}
for $z\to \infty$ and $0<\gamma<2/3$, 
where the constants $S_{0}$ and $W_{0}$ may be different from 
the exact flat Friedmann case.
Equation~(\ref{speed}) implies
\begin{equation}
V\approx V_{0}z^{1-2/(3\gamma)},\quad V_{0}=\frac{c_{1}}{c_{0}}S_{0}^{-2}W_{0}^{(\gamma-2)/\gamma}.
\end{equation}
The exact Friedmann relation (\ref{eq:relation2_Friedmann})
still applies but Eq.~(\ref{00+11}) now yields to lowest order
\begin{equation}
\frac{2}{3\gamma}V_{0}^{2}+\frac{1-\gamma}{\gamma}
\left(\frac{W'}{W}-2\right)z^{-2(1-2/(3\gamma))}=\frac{\gamma
 c_{1}^{2}W_{0}^{(\gamma-2)/\gamma}}
{2 S_{0}^{4}}.
\end{equation}
Hence $(W'/W)-2$ is either proportional to $z^{2-4/(3\gamma)}$ or falls
off even faster. Therefore Eq.~(\ref{01}) implies the following asymptotic behavior:
\begin{eqnarray}
S&=&S_{0}z^{-2/(3\gamma)}\left[1+S_{1}z^{2(1-2/(3\gamma))}+\cdots\right], \\
W&=&W_{0}z^{2}\left[1+W_{1}z^{2(1-2/(3\gamma))}+\cdots\right],
\end{eqnarray}
where 
\begin{eqnarray}
&&\frac{2}{3\gamma}V_{0}^{2}+\frac{1-\gamma}{\gamma} W_{1}=\frac{\gamma
 c_{1}^{2}W_{0}^{(\gamma-2)/\gamma}}{2 S_{0}^{4}}, \\
&&S_{1}=-\frac{1}{5\gamma(3-2\gamma)}W_{1}.
\end{eqnarray}
All coefficients of higher order terms are determined from $S_{0}$ and
$W_{0}$. Since the value for $S_{0}$ can be different
from the exact Friedmann case, there is a 1-parameter family of 
asymptotically quasi-Friedmann solutions at small distances.
Since $V\to 0 $ as $z\to \infty$, $z=\infty$
is a vertical line in a conformal diagram.
These solutions can be interpreted as self-similar models with 
a regular center as $r\to 0$ for fixed $t\ne 0$.
On the other hand, in the limit $t\to 0$ for fixed $r\ne 0$,
they correspond to simultaneous big bang models in the comoving time-slicing.
If $S_{0}$ had the exact Friedmann value, we would
have Friedmann itself, so this is the only solution which is properly asymptotically Friedmann.

\subsection{Static asymptote for $z\to \infty$}

As we have seen, there exists no static self-similar 
solution for $0<\gamma<2/3$.
However, we will show that there are still solutions which can be described as {\it asymptotically quasi-static}.
In this case, we assume 
\[
W\approx W_{0}>0, \quad S\approx S_{0}>0
\]
as $z\to \infty$. Only if $S_{0}$ and $W_{0}$ are the same as in the static solution do we describe such solutions as {\it asymptotically static}.
Equation~(\ref{speed}) implies 
\begin{equation}
V\approx V_{0}z^{(2-\gamma)/\gamma}, \quad V_{0}=\frac{c_{1}}{c_{0}}z^{(2-\gamma)/\gamma}S_{0}^{-2}W_{0}^{(\gamma-2)/\gamma}.
\end{equation}
This solution can be compatible with Eq.~(\ref{00+11}) if and only if 
\begin{equation}
\frac{1}{\gamma}\frac{W'}{W}+2\frac{S'}{S}\propto z^{-2(2-\gamma)/\gamma},
\label{eq:HOT_combination}
\end{equation}
as $z\to \infty$.
This combination appears in the second term 
on the right-hand side of Eq.~(\ref{01}) 
and this can be regarded as a higher order term.
It then follows that
\begin{equation}
S=S_{0}+S_{1}z^{1-2/\gamma}+S_{2}z^{2(1-2/\gamma)}+\cdots.
\end{equation}
From Eq.~(\ref{eq:HOT_combination}), $W$ must be of the form
\begin{equation}
W=W_{0}+W_{1}z^{1-2/\gamma}+W_{2}z^{2(1-2/\gamma)}+\cdots,
\end{equation}
where 
\begin{equation}
\frac{1}{\gamma}\frac{W_{1}}{W_{0}}+2\frac{S_{1}}{S_{0}}=0.
\label{eq:AQS_1st_order_1}
\end{equation}
From Eq.~(\ref{eq:constraint}), we obtain the relation
\begin{equation}
W_{0}S_{0}^{3}=S_{0}\left[1+\left(1-\frac{2}{\gamma}\right)^{2}
c_{0}^{-2}W_{0}^{2(1-2/\gamma)}S_{1}^{2}
-c_{1}^{-2}S_{0}^{6}W_{0}^{2/\gamma}\right].
\label{eq:AQS_1st_order_2}
\end{equation}
To lowest order Eq.~(\ref{00+11}) becomes 
\begin{equation}
2V_{0}^{2}\left(1-\frac{2}{\gamma}\right)
\left[\frac{1}{\gamma}\frac{W_{2}}{W_{0}}
+2\frac{S_{2}}{S_{0}}\right]=\frac{\gamma c_{1}^{2}W_{0}^{1-2/\gamma}}
{2S_{0}^{4}}+\frac{2(1-\gamma)}{\gamma}
\label{eq:AQS_2nd_order_1}
\end{equation}
and to the next lowest order Eq.~(\ref{01}) becomes
\begin{eqnarray}
2\left(1-\frac{2}{\gamma}\right)^{2}S_{2}&=&
-\left(1-\frac{1}{\gamma}\right)\left(1-\frac{2}{\gamma}\right)^{2}S_{1}W_{1}
\nonumber \\
&-& 2\left(1-\frac{1}{\gamma}\right)S_{0}\left(
\frac{2S_{2}}{S_{0}}+\frac{W_{2}}{\gamma W_{0}}\right).
\label{eq:AQS_2nd_order_2}
\end{eqnarray}
$S_{1}$ and $W_{1}$ are determined from Eqs.~(\ref{eq:AQS_1st_order_1})
and (\ref{eq:AQS_1st_order_2}) in terms of $S_{0}$ and $W_{0}$, while 
$S_{2}$ and $W_{2}$ are determined by Eqs.~(\ref{eq:AQS_2nd_order_1})
and (\ref{eq:AQS_2nd_order_2}) in terms of $S_{0}$, $S_{1}$, $W_{0}$ 
and $W_{1}$.
All higher order terms are determined in terms
of $S_{0}$ and $W_{0}$. Hence there is a 2-parameter family of 
asymptotically quasi-static solutions. Since there is no exact
static solution, there are no asymptotically static ones.
Also $V\to \infty $ as $z\to \infty$ for $0<\gamma<2/3$, so 
$z=\infty$ is a horizontal line in 
a conformal diagram.

If we consider the analytic continuation beyond $z=\infty$ into the negative $z$ region, 
the metric should remain analytic in terms of 
the local inertial Cartesian coordinates.
For asymptotically quasi-static solutions, one should consider 
the analytic continuation beyond $t=+0$ because $S=R/r$ and $W=8\pi G
\mu r^2$ are finite. For this purpose,
it is convenient to see how the proper time $\tau$ 
behaves for fixed $r>0$. For this class of solutions, Eq.~(\ref{integral1}) implies
\begin{equation}
e^{\Phi}=c_{0}z^{-2(1-\gamma)/\gamma}W_{0}^{(1-\gamma)/\gamma},
\end{equation}
so $\tau\propto t^{-1+2/\gamma}$. When we continue
the solution analytically to the negative $t$ region beyond $t=\pm 0$, 
this means that 
$\tau\approx \pm C_{r}|t|^{-1+2/\gamma}$, where $C_{r}$
is a positive constant depending on $r$ and the upper (lower)
sign corresponds to the positive (negative) $t$.
To express $W$ and $S$ as analytic functions of $\tau$, 
we use the following unique continuation:
\begin{eqnarray}
S&=&S_{0}\pm S_{1}|z|^{1-2/\gamma}+S_{2}|z|^{2(1-2/\gamma)}+\cdots, \label{asympQS-S} \\
W&=&W_{0}\pm W_{1}|z|^{1-2/\gamma}+W_{2}|z|^{2(1-2/\gamma)}+\cdots. \label{asympQS-W}
\end{eqnarray}
This expression gives a Taylor series expansion 
around $\tau=0$ in terms of $\tau$. 
Because of the presence of the odd powers of $\tau$,
this is not reflection-symmetric about $t=0$.

On the other hand, in the limit $r\to \infty$ for fixed $t\ne 0$,
one has a vacuum of infinite radius which is not asymptotically
flat because $m/R$ approaches a non-zero constant. We call this
{\it quasi-static spacelike infinity.}
In Paper II, we will see that there are a class of 
solutions describing a Friedmann universe emergent from a white hole,
where a Friedmann spacelike infinity and a quasi-static spacelike 
infinity are connected.

\subsection{Kantowski-Sachs asymptote for $z\to \infty $}
For the {\it asymptotically quasi-Kantowski-Sachs} solutions, we assume
\begin{equation}
S\approx S_{0}z^{-1},\quad W\approx W_{0}z^{2}
\end{equation}
as $z\to \infty$. 
From Eq.~(\ref{speed}), we have
\begin{equation}
V\approx V_{0}z^{3-2/\gamma}, \quad V_{0}=\frac{c_{1}}{c_{0}}S_{0}^{-2}
W_{0}^{1-2/\gamma}z^{3-2/\gamma},
\end{equation}
and hence $V\to 0$ as $z\to \infty$.
Equation~(\ref{00+11}) is trivially satisfied
to lowest order 
but to the next order requires either
\begin{equation}
\frac{W'}{W}-2 \propto z^{2(3-2/\gamma)}
\label{eq:HOT_combination_2}
\end{equation}
or that the right-hand side falls off even faster than this.
Equations~(\ref{01}) and~(\ref{eq:HOT_combination_2}) imply
\begin{eqnarray}
S&=&S_{0}z^{-1}(1+S_{1}z^{3-2/\gamma}+S_{2}z^{2(3-2/\gamma)}+\cdots), \\
W&=&W_{0}z^{2}(1+W_{2}z^{2(3-2/\gamma)}+\cdots)
\end{eqnarray}
Equation~(\ref{eq:constraint}) yields
\begin{eqnarray}
(1-\gamma)W_{0}S_{0}^{3}&=&S_{0}\left[1+c_{0}^{-2}W_{0}^{2(\gamma-1)/\gamma}S_{0}^{2}
\right.\nonumber \\
& &\left.-\left(3-\frac{2}{\gamma}\right)^{2}c_{1}^{-2}S_{0}^{6}W_{0}^{2/\gamma}S_{1}^{2}\right],
\end{eqnarray}
while Eq.~(\ref{00+11}) gives
\begin{equation}
\frac{2(1-\gamma)}{\gamma}V_{0}^{2}+2\frac{1-\gamma}{\gamma}
\left(3-\frac{2}{\gamma}\right)W_{2}
=\frac{\gamma c_{1}^{2}W_{0}^{1-2/\gamma}}{2S_{0}^{4}}.
\label{eq:next_order_2.24}
\end{equation}
At second order Eq.~(\ref{01}) becomes 
\begin{equation}
\left(4-\frac{3}{\gamma}\right)S_{2}=-\left(1-\frac{1}{\gamma}\right)
\left(3-\frac{2}{\gamma}\right)\left(W_{2}+S_{1}^{2}\right).
\label{eq:AKS_2nd_order_2}
\end{equation}
Hence the coefficients of all higher order terms are determined in
terms of $S_{0}$ and $W_{0}$. This means there is a 2-parameter family of 
asymptotically quasi-Kantowski-Sachs solutions.
If the values for $S_{0}$ and $W_{0}$ are 
the same as those for the exact Kantowski-Sachs solution, all
coefficients of higher order terms vanish, so the solution is 
also exactly Kantowski-Sachs.
In other words, the only solution which is 
asymptotically Kantowski-Sachs is Kantowski-Sachs itself.
On the other hand, even if the first order terms 
vanish, i.e. $S_{1}=0$, we can see that $W_{2}$ may not vanish, so $S_{0}$
and $W_{0}$ are different from their Kantowski-Sachs values.
Since $V\to 0 $ as $z\to \infty$ for $0<\gamma<2/3$, 
$z=\infty$ is a vertical line in 
a conformal diagram.

For asymptotically quasi-Kantowski-Sachs solutions, one can consider 
the analytic continuation beyond $r=\infty$ because $zS=R/t$ and 
$W/z^{2}=8\pi G\mu t^{2}$ are non-zero and finite for fixed $t\ne 0$. 
It is useful to see how the proper length $\lambda$ 
changes around $r=\infty$ for fixed $t\ne 0$.
Equation~(\ref{integral2}) implies
\begin{equation}
e^{\Psi}=c_{1}z^{2-2/\gamma}S_{0}^{-2}W_{0}^{-1/\gamma},
\end{equation}
so $\lambda \propto r^{3-2/\gamma}$. This means that, when we continue
the solution analytically to the negative $r$ region beyond $r=\infty$, one has $\lambda\approx \mp C_{t}|r|^{3-2/\gamma}$, 
where $C_{t}$ is a positive constant depending on $t$ and 
the upper (lower) sign corresponds to positive (negative) $r$.
To obtain analytic expressions for $W$ and $S$ in terms of $\lambda$, 
we use the following continuation:
\begin{eqnarray}
S&=&S_{0}z^{-1}(1 \pm S_{1}|z|^{3-2/\gamma}+S_{2}|z|^{2(3-2/\gamma)}+\cdots), \label{asympQKS-S}\\
W&=&W_{0}z^{2}(1 +W_{2}|z|^{2(3-2/\gamma)}+\cdots). \label{asympQKS-W}
\end{eqnarray}
The above expression gives a Taylor series expansion 
around $\lambda=0$.
Since $W_{1}=0$ in the above expression, $W/z^{2}=8\pi G t^{2}$ 
generically has an extremum at $z=\pm \infty$, while $zS=R/t$ does not.
So this expression is not reflection-symmetric with respect to $r=\pm\infty$.
On the other hand, in the limit $t\to 0$
for fixed $r\ne 0$, we have $R\to 0$ and $\mu\to \infty$, which corresponds to
an initial singularity. In Paper II, we will see that this analytical
continuation 
 is crucial
for obtaining black hole solutions embedded in a Friedmann background
as well as a class of wormhole solutions connecting one Friedmann 
spacelike infinity to another quasi-Friedmann spacelike infinity.

\subsection{Constant-velocity asymptote for $z\to \infty$}
Here we seek solutions 
in which $V$ tends to a finite positive value $V_{\infty}$ 
as $z\to \infty$.
Differentiating Eq.~(\ref{speed}) and noting that $V'/V \to 0$ as $z\to \infty$, we 
have 
\begin{equation}
\frac{W'}{W}\approx 1-\left( \frac{2\gamma}{2-\gamma}\right) \frac{S'}{S}.
\end{equation}
Substituting this relation into Eq.~(\ref{01}), we find to lowest order
\begin{equation}
\frac{S''}{S}\approx -\frac{1}{\gamma}-\frac{4(1-\gamma)}{(2-\gamma)}
\left(\frac{S'}{S}\right)^{2}-\frac{2(2-\gamma^{2})}{\gamma(2-\gamma)}
\frac{S'}{S}.
\end{equation}
To lowest order this ordinary differential equation has a general solution
of the form
\begin{equation}
S\approx S_0^+z^{p_{+}}+S_0^-z^{p_{-}},
\label{eq:V*_S_possibility}
\end{equation}
where $S_{0,\pm}$ are arbitrary constants and
\begin{equation}
p_{\pm}=\frac{-(2-\gamma^{2})\pm \sqrt{(1-\gamma)(4-8\gamma+4\gamma^{2}-\gamma^{3})}}{(6-5\gamma)\gamma}.
\end{equation}
One can show that the square root is real
for $0<\gamma<2/3$. Also
Eq.~(\ref{eq:constraint}) is consistent only 
if the first term dominates the second term in 
Eq.~(\ref{eq:V*_S_possibility}),
so $S_0^+=0$ is excluded. 

\begin{figure}
\begin{center}
\includegraphics[width=0.8\linewidth]{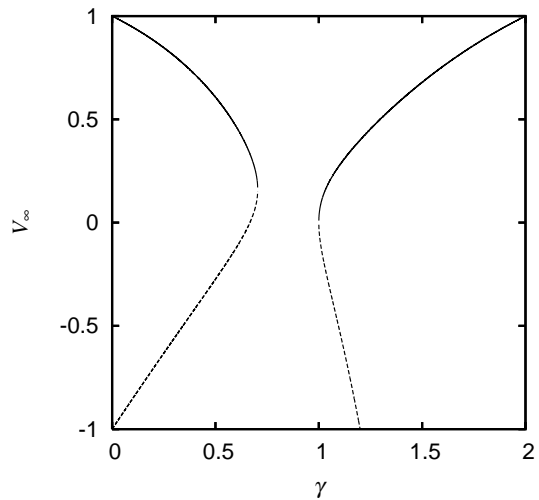}
\caption{\label{fg:v_inf}$V_{\infty}$ for the asymptotically 
constant-velocity solution as a function of $\gamma$.
Although only the part of the solid curve with $0<\gamma<2/3$ is 
relevant to the present analysis, it is extended to $0< \gamma < 2$ 
for completeness. The other root for $V_{\infty}$ is also 
plotted with a dashed line but this may be unphysical.
}
\end{center}
\end{figure}

One can easily 
show that the first term on the right-hand side 
of Eq.~(\ref{00+11}) converges to zero for 
$0<\gamma<2/3$ and so
\begin{equation}
 V_{\infty}=\frac{\gamma(1-\gamma)+\sqrt{(1-\gamma)(4-8\gamma+4\gamma^{2}-\gamma^{3})}}{2-\gamma}.
\label{eq:V_star}
\end{equation}
This is plotted in Fig.~\ref{fg:v_inf}.
Note that the solution with similar asymptotic behavior for $\gamma>6/5$
~\cite{cc2000a} has 
 the other branch of the square root. 
For this solution, we have
\begin{equation}
S\approx S_{0}z^{p_{+}}, \quad W\approx W_{0}z^{1-2\gamma p_{+}/(2-\gamma)}.
\end{equation}
The condition $V\to V_{\infty}$
yields a relation between $S_{0}$ and $W_{0}$:
\begin{equation}
V_{\infty}=\frac{c_{1}}{c_{0}}S_{0}^{-2}W_{0}^{(\gamma-2)/\gamma}.
\end{equation}
There is  no other relation between
them, so there is a 1-parameter family of solutions
with this asymptotic behavior.
Since $V\to V_{\infty}<1 $ as $z\to \infty$ for $0<\gamma<2/3$, 
$z=\infty$ is a vertical line in 
a conformal diagram.

We now discuss the physical significance of
this asymptotic solution.
As $\gamma$ increases from 0 to 2/3, $p_{+}$ decreases from $-1/2$ to
$-3/4$, while $V_{\infty}$ decreases from 1 to 1/3.
Because $M\propto z^{1+(6-5\gamma)p_{+}/(2-\gamma)}$, we have
$R=rS \to \infty$, $\mu=W/(8\pi G r^{2}) \to 0$, 
$m=rM/(2G)\to \infty$ and $2m/R=M/S\to \infty$
as $r\to \infty $ for fixed $t\ne 0$.
Therefore, this solution approaches a vacuum region 
with infinite physical radius.
However, because the fall-off of the energy density $\mu$ is very slow,
the solution is far from asymptotically flat.
In fact, both $m$ and $m/R$ diverge to 
infinity.
This solution cannot be analytically extended beyond $z=\infty$ 
because $R$ diverges to infinity as $r\to \infty $ for fixed $t\ne 0$,
while $\mu$ diverges to infinity as $t\to 0$ for fixed $r\ne 0$.
We describe these as {\it asymptotically constant-velocity} solutions.
The infinity reached as $r\to \infty$ for fixed $t\ne 0$ will be called the
{\it constant-velocity spacelike infinity} (cf. the asymptotically Minkowski
solutions for $\gamma > 6/5$ described in CC).

\subsection{Singular asymptote for $z\to z_{*}$}
We assume that $\ln |S|$ diverges as $z\to z_{*}$, 
while $V$ tends to a finite value $V_{*} $. 
Then Eq.~(\ref{speed}) implies 
\begin{equation}
\frac{W'}{W}\approx -\frac{2\gamma}{2-\gamma}\frac{S'}{S}
\end{equation}
and Eq. (\ref{01}) yields to lowest order
\begin{eqnarray}
S&\approx &S_{0}|Z|^{(2-\gamma)/(6-5\gamma)}, \\
W&\approx &W_{0}|Z|^{-2\gamma/(6-5\gamma)},
\end{eqnarray}
where 
\begin{equation}
Z\equiv\ln \frac{z}{z_{*}}.
\end{equation}
Equation~(\ref{00+11}) implies $V_{*}^{2}=1$, while
Eq.~(\ref{m1}) gives
\begin{equation}
M\to M_{0}=\pm \frac{\gamma(2-\gamma)}{6-5\gamma}W_{0}S_{0}^{3},
\label{eq:singular_M0}
\end{equation}
where the upper (lower) sign corresponds to positive (negative) $Z$.
From Eq.~(\ref{speed}), we have
\begin{equation}
\frac{c_{1}^{2}}{c_{0}^{2}}|z_{*}|^{2(2-\gamma)/\gamma}S_{0}^{-4}W_{0}
^{2(\gamma-2)/\gamma}=1
\end{equation}
and hence $|z_{*}|$ is determined from $S_{0}$ and $W_{0}$. 
To next lowest order, Eq.~(\ref{01}) implies 
\begin{eqnarray}
S=S_{0}|Z|^{(2-\gamma)/(6-5\gamma)}
\left[1+
S_{1}|Z|^{(2-3\gamma)/(6-5\gamma)}+\cdots
\right], \\
W=W_{0}|Z|^{-2\gamma/(6-5\gamma)}
\left[1+
W_{1}|Z|^{(2-3\gamma)/(6-5\gamma)}+\cdots
\right],
\end{eqnarray}
where
\begin{equation}
S_{1}=-\frac{(2-\gamma)}{2(4-3\gamma)}W_{1}.
\end{equation}
Through Eq.~(\ref{speed}), we obtain
\begin{equation}
V=1+V_{1}|Z|^{(2-3\gamma)/(6-5\gamma)}+\cdots, \label{asympSing-V}
\end{equation}
where
\begin{equation}
V_{1}=-2S_{1}-\left( \frac{2-\gamma}{\gamma}\right) W_{1},
\end{equation}
and hence
\begin{eqnarray}
S_{1}&=&\frac{\gamma}{8(1-\gamma)}V_{1}, \\
W_{1}&=&-\frac{\gamma(4-3\gamma)}{4(2-\gamma)(1-\gamma)}V_{1}.
\end{eqnarray}
To next lowest order Eq.~(\ref{00+11}) then gives 
\begin{equation}
V_{1}=\pm \left( \frac{6-5\gamma}{2-\gamma}\right)
\frac{\gamma c_{1}^{2}W_{0}^{-(2-\gamma)/\gamma}}{2S_{0}^{4}},
\end{equation}
which is consistent with Eq.~(\ref{eq:constraint}). 

We describe these solutions as {\it asymptotically singular}.
For fixed $r\ne 0$, we have 
$R=rS\to 0$, $\mu =W/(8\pi Gr^{2})\to \infty$ and $m=r M/(2G) 
\approx rM_{0}/(2G)$ as $Z\to 0$.
From Eq.~(\ref{eq:singular_M0}), we have 
\begin{equation}
 m\approx \frac{rM_{0}}{2G}=\pm \frac{\gamma(2-\gamma)}{6-5\gamma}W_{0}S_{0}^{2}R
\end{equation}
and so the mass is positive (negative) for $Z\to 0^+$ ($Z\to 0^-$).
(Note that $S$ can be negative but $R=rS$ is positive
even after the extension to the negative $r$ region.)
Since the physical properties of these two limits are very different, 
we distinguish these two cases as
{\it asymptotically positive-mass singular}
and {\it asymptotically negative-mass singular}.
For the positive-mass singular case,
$Z \to 0^+$, $V\to 1^+$ and $m>0$; for the negative case,
$Z \to 0^-$, $V\to 1^-$ and $m<0$. 
Although $V\to 1$ in the limit $Z\to 0$, the first component of the metric~(\ref{metric2}) is
\begin{eqnarray}
&& -(1-V^{2})e^{2\Phi}  \nonumber \\
&\approx& 2c_{0}^{2}|z_{*}|^{4(\gamma-1)/\gamma}
W_{0}^{2(1-\gamma)/\gamma}V_{1}|Z|^{-(2-\gamma)/(6-5\gamma)},
\end{eqnarray}
which tends to $\pm\infty$. Hence the similarity surface becomes spacelike (timelike)
for the positive (negative) mass case.
We will see in Paper II that the positive-mass singular 
behavior is associated with black hole
and white hole singularities,
while 
the negative-mass one is associated with 
naked singularities.

It should be noted, however, that the mass of the singularity is not
constant but is proportional to the time coordinate $t$.
There is a 2-parameter family of solutions belonging to this class.
 
\section{Summary and discussion} 
In this paper, we have analyzed spherically symmetric 
self-similar spacetimes. 
We have shown that the metric of these spacetimes is conformally static,
with the conformal structure being determined by the form of the velocity 
function $V$. The homothetic Killing vector here plays an analogous role to
the Killing vector in the static case. We have expressed the Einstein field equations in the comoving 
approach as ordinary 
differential equations and analytically investigated the possible asymptotic 
behavior of
perfect fluid solutions with
$p=(\gamma-1)\mu$ and $0<\gamma<2/3$. In Paper II, we focus on solutions which are asymptotic 
to the flat Friedmann universe at large distances because we are interested in solutions embedded in an
accelerating Friedmann background~\cite{mhc2}. 

These solutions can be understood in terms of the ``complete classification''
provided by Carr and Coley~\cite{cc2000a}.  
However, this classification mainly applies for
perfect fluids with positive pressure and we have extended
it to $0<\gamma<2/3$.
We find there are eight possible asymptotic 
behaviors at small, intermediate and large distances 
in this class of solutions.
However, it should be stressed that there are other asymptotic behaviors, 
such as Kantowski-Sachs asymptotes at $z \to \pm 0$, and these are discussed elsewhere~\cite{chm2007}.

Tables~\ref{table:exact_solutions} and \ref{table:asymptotic_behaviors}
summarize the results.
Table~\ref{table:exact_solutions} 
lists two exact solutions: 
Friedmann (F) and Kantowski-Sachs (KS).
Table~\ref{table:asymptotic_behaviors} 
lists eight asymptotic behaviors:
asymptotically Friedmann (F) and asymptotically quasi-Friedmann (QF)
for $z\to 0$,
asymptotically quasi-Friedmann (QF), 
asymptotically quasi-static (QS),
asymptotically quasi-Kantowski-Sachs (QKS) and
asymptotically constant-velocity (CV) 
for $z\to \infty$,
asymptotically positive-mass singular (PMS) and asymptotically
negative-mass singular (NMS) for $z\to z_{*}$.

\begin{table}[htbp]
\begin{center}
\caption{\label{table:exact_solutions}Self-similar exact solutions for
 $0<\gamma<2/3$. The form
of $S$, $W$ and $V$, the number of parameters, the
analytic continuation and the causal structure are shown.
There is no arbitrary parameter in these solutions.
}
\begin{tabular}{|c||c|c|c|c|c|}
\hline
Name & $S$ & $W$ & $V$ & Continuation & Structure\\ \hline \hline
F & $ z^{-\frac{2}{3\gamma}}$ & $ z^{2} $ &$ z^{1-\frac{2}{3\gamma}}$ &
None & Fig.~\ref{fig:Friedmann}\\ \hline
KS & $ z^{-1}$ & $ z^{2}$ &  $ z^{3-\frac{2}{\gamma}}$ & $r=\pm \infty$
 & Fig.~\ref{fig:KS}\\ \hline 
\end{tabular}
\end{center}
\end{table}

\begin{table*}
\begin{center}
\caption{\label{table:asymptotic_behaviors}
The possible asymptotic behaviors of asymptotically Friedmann 
self-similar solutions for $0<\gamma<2/3$. 
The limiting value of $z$ and the form
of $S$, $W$ and $V$, 
the number of parameters, the
analytic continuation, the causal structure
and the physical distance are shown.
For the physical distance, the limiting values
of $zS=R/t$ are shown.
The values of $p_{+}=p_{+}(\gamma)$ and $V_{\infty}=V_{\infty}(\gamma)$ are given in the text.}
\begin{tabular}{|c||c|c|c|c|c|c|c|c|}
\hline
Name  & z & $S$ & $W$ &$V$ & \#param & Extension & Structure & Distance \\ \hline \hline
F & $\pm 0$ & $ z^{-\frac{2}{3\gamma}}$ 
& $ z^{2} $ & $ z^{1-\frac{2}{3\gamma}}$  & 1 & None & Spacelike & $\infty$ \\ \hline
QF & $\pm 0$ & $ z^{-\frac{2}{3\gamma}}$ & $ z^{2}$ & $
 z^{1-\frac{2}{3\gamma}}$ & 1 & None & Spacelike & $\infty$ \\ \hline
QF & $\pm \infty$ & $ z^{-\frac{2}{3\gamma}}$ & $ z^{2}$ & $
 z^{1-\frac{2}{3\gamma}}$ & 1 & None & Timelike & 0 \\ \hline
QS & $\pm \infty$ & const & const & $ z^{\frac{2-\gamma}{\gamma}}$ 
& 2 & $t=\pm 0$ & Spacelike & $\infty$ \\ \hline
QKS & $\pm \infty$  & $ z^{-1}$  & $ z^{2}$ &$
 z^{3-\frac{2}{\gamma}}$  & 2  & $r=\pm \infty$ & Timelike & Intermediate \\ \hline
CV & $\pm \infty$  & $ z^{p_{+}}$& $ z^{1-\frac{2\gamma
 p_{+}}{2-\gamma}}$ & $V_{\infty}$ & 1 & None & Timelike & $\infty$ \\ \hline
PMS & $z_{*}$ & $ |z-z_{*}|^{\frac{2-\gamma}{6-5\gamma}}$ &
 $|z-z_{*}|^{-\frac{2\gamma}{6-5\gamma}}$& $1$ & 2 & None & Spacelike & 0 \\ \hline
NMS & $z_{*}$ & $ |z-z_{*}|^{\frac{2-\gamma}{6-5\gamma}}$ &
  $|z-z_{*}|^{-\frac{2\gamma}{6-5\gamma}}$ & $1$ & 2 & None & Timelike & 0\\ 
\hline
\end{tabular}
\end{center}
\end{table*} 

There are three key differences from the positive pressure case.
First, there is no sonic point in the 
negative-pressure case and hence no additional ambiguity associated with the sonic point.
Second, these solutions can be 
properly asymptotic to the flat Friedmann model, 
in the sense that there is no solid angle deficit.
This contrasts with the positive pressure case, where
the solutions are only asymptotically
quasi-Friedmann.
Third, the existence 
of exact and asymptotically Kantowski-Sachs solutions,
which can be extended
beyond the timelike hypersurface $z=\pm \infty$ along the spacelike
direction, leads to physically interesting
solutions embedded in a Friedmann background. 
The analytical results of this paper 
play a crucial role in Paper II, where we investigate  the new solutions numerically 
and interpret them physically. 

Finally, it should be emphasized that the 
matter model considered here suffers from
dynamical
instability for small-scale perturbation. 
For a perfect fluid with equation of state $p=(\gamma-1)\mu$,
the derivative $dp/d\mu=\gamma-1$ is negative for $\gamma<1$, which
violates the condition for 
dynamically
stable equilibrium.
We can also discuss this instability in terms of 
linear perturbation analysis.
Consider a density perturbation with the dependence
$e^{i(\omega t-\mbox{\boldmath $k$}\cdot \mbox{\boldmath $x$})}$ to
linear order in flat spacetime.
Then the dispersion relation $\omega^{2}=(\gamma-1)\mbox{\boldmath
$k$}^{2}$ is obtained. Since the wave-number $\mbox{\boldmath $k$}$
is real, $\omega=\pm i \sqrt{1-\gamma} |\mbox{\boldmath $k$}|$ for
$\gamma<1$, the lower sign corresponding to exponential growth
of the perturbation. 
Since this instability exists for 
arbitrarily short wavelengths, the system is unstable even
in the presence of gravity, except for the case of a cosmological constant
($\gamma=0$). This means that the matter fields discussed in this paper 
can only be valid 
(i.e. dynamically stable) 
at larger scales.
In fact, there are important physical models
which have an effectively
negative pressure through 
coarse-graining but which are still absolutely stable at small scales.
For example, the effective equations of state for domain wall networks and string
networks correspond to $\gamma=1/3$ and 
$\gamma=2/3$, respectively~\cite{vs1994}.

It should also be noted that this small-scale 
dynamical 
instability does not appear for a quintessence field (i.e. a scalar
field with a flat potential).
For example, suppose the scalar potential is given by
$V=V_{0}\exp (\sqrt{8\pi} \lambda \phi)$ ($\lambda>0$).
This can accelerate the Friedmann universe for $\lambda^{2}<2$.
The dispersion relation for the scalar field perturbation 
in flat spacetime becomes $\omega^{2}=\mbox{\boldmath $k$}^{2}+8\pi \lambda^2 V(\phi_{0})$,
where $\phi_0(t,\mbox{\boldmath $x$})$
denotes the background scalar field and the short wavelength
approximation is adopted. Thus, this scalar field is stable
against short wavelength perturbations.

\acknowledgments
The authors would like to thank P.~Ivanov, H.~Kodama, H.~Koyama, M.~Siino and T.~Tanaka for useful comments. 
HM and TH are supported by the Grant-in-Aid for Scientific
Research Fund of the Ministry of Education, Culture, Sports, Science
and Technology, Japan (Young Scientists (B) 18740162 (HM) and 18740144
(TH)). 
HM was also supported by the Grant No. 1071125 from FONDECYT (Chile).
CECS is funded in part by an institutional grant from Millennium Science
Initiative, Chile, and the generous support to CECS from Empresas CMPC 
is gratefully acknowledged.
BJC thanks the Research Center for the Early Universe at the University of Tokyo for hospitality received during this work.

\appendix

\section{Comparison with Carr-Coley analysis.}

Carr and Coley~\cite{cc2000b} have also attempted to classify perfect-fluid spherically symmetric self-similar solutions with $0<\gamma<1$, although they have not investigated the solutions numerically (as we do in Paper II). Their analysis essentially extends the CC classification~\cite{cc2000a} of solutions with  $1<\gamma<2$ and, in principle, includes the $0<\gamma<2/3$ case discussed here. However, a comparison with our results requires care because CC use a different system of notation from us.  This appendix gives the relationship between the two systems and points out that the analysis in ref.~\cite{cc2000b} is not complete. 

CC use the same comoving coordinates ($r, t,\theta,\phi$) as us and the same similarity variable $z=r/t$, although a dot rather than a prime denotes $d/d\ln z$. They also use the same symbols for the velocity function $V$, pressure $p$, density $\mu$ and areal radial distance $R=rS$. Otherwise the mapping from our variables to theirs is as follows:
\begin{eqnarray}
\Phi \rightarrow \nu, \quad \Psi \rightarrow \lambda, \quad M \rightarrow MS, \\
\quad P \rightarrow 8\pi G P/S^2, \quad W \rightarrow 8\pi G W/S^2.
\end{eqnarray}
They also express the field equations as ODEs in the quantities ($S, S', W$) but they introduce the variable
\begin{equation}
 x \equiv (4\pi \mu r^2)^{-(1+\gamma)/(2+\gamma)}
\end{equation}
instead of $W$. Note that the ODEs are expressed in terms of ($S,M,W$) in ref.~\cite{hm2002}.

CC correctly identify Friedmann and Kantowski-Sachs as the only exact power-law self-similar solutions and they look for more general solutions which are asymptotic to these. They also point out that there is no exact static solutions for $0<\gamma <1$ but they do not notice that there are still asymptotically quasi-static solutions and this is the main reason why their classification is incomplete. Although CC appreciate the importance of extending some solutions from $z>0$ to $z<0$, they always take the variable $r$ to be positive, whereas we allow it to be negative. 

Their analysis of the asymptotic solutions at large and small distances is closely related to ours and explicitly includes the asymptotically Friedmann and asymptotically Kantowski-Sachs solutions for $0<\gamma<2/3$. However, they do not distinguish between the asymptotically Friedmann and asymptotically quasi-Friedmann solutions. Their description of the asymptotic-Kantowski-Sachs solutions is also somewhat different: because $r$ is taken to be positive, they do not extend this solution from $z>0$ to $z<0$, so they do not notice these solutions are related to wormholes (cf. Paper II). 

CC derive the asymptotically constant-velocity solutions, as well as deriving and plotting Eq.~(\ref{eq:V_star}), but they describe these as ``asymptotically Minkowski'' and do fully understand their physical interpretation. They also identify the asymptotic solutions which represent a central singularity  (i.e. in which $V \rightarrow 1$ at finite $z$) but they only include them for $2/3<\gamma<1$ and omit them for $0<\gamma<2/3$. They also make no distinction between the positive and negative mass singularities.  They claim to find another small-distance solution for $2/3<\gamma<1$, which they label ``Y'',  but we do not find this and it is probably non-physical. 

Analogue of Tables I and II appears in ref.~\cite{cc2000b} but no
analysis of the global structure of the solutions is included, which is
another reason why the wormhole connection is missed. 
Carr and Gundlach~\cite{cg2003} have analyzed the global structure of
some of these solutions but there is a mistake in the asymptotically
Kantowski-Sachs solutions shown in Fig. 15, where 
$z=\infty$ is wrongly identified with the singularity.

\end{document}